\documentclass[aps,prd,preprint,superscriptaddress]{revtex4-2}

\usepackage{ulem}
\usepackage{xcolor}
\usepackage[colorlinks=true]{hyperref}
\usepackage{soul}
\usepackage{booktabs}
\usepackage{mathrsfs}
\usepackage{amssymb}

\usepackage{amsmath,amssymb,color,epsfig}
\allowdisplaybreaks[4]

\newcommand{\be}{\begin{equation}}
\newcommand{\ee}{\end{equation}}
\newcommand{\bea}{\setlength\arraycolsep{2pt} \begin{eqnarray}}
\newcommand{\eea}{\end{eqnarray}}
\newcommand{\nn}{\nonumber}

\def\0{{\sst{(0)}}}
\def\1{{\sst{(1)}}}
\def\2{{\sst{(2)}}}
\def\3{{\sst{(3)}}}
\def\4{{\sst{(4)}}}
\def\5{{\sst{(5)}}}
\def\6{{\sst{(6)}}}
\def\7{{\sst{(7)}}}
\def\8{{\sst{(8)}}}
\def\sst#1{{\scriptscriptstyle #1}}

\begin{document}

\hypersetup{
    linkcolor=blue,
    citecolor=red,
    urlcolor=magenta
}


\title{Black holes and neutron stars in massive Hellings-Nordtvedt theory}



\author{Zhe Luo}
\affiliation{Department of Physics, Key Laboratory of Low Dimensional Quantum Structures and Quantum Control of Ministry of Education, and Institute of Interdisciplinary Studies, Hunan Normal University, Changsha, 410081, China}
\affiliation{Hunan Research Center of the Basic Discipline for Quantum Effects and Quantum Technologies, Hunan Normal University, Changsha 410081, China}

\author{Liang Liang}
\affiliation{Department of Physics, Key Laboratory of Low Dimensional Quantum Structures and Quantum Control of Ministry of Education, and Institute of Interdisciplinary Studies, Hunan Normal University, Changsha, 410081, China}
\affiliation{Hunan Research Center of the Basic Discipline for Quantum Effects and Quantum Technologies, Hunan Normal University, Changsha 410081, China}

\author{Zhong-Xi Yu}
\email[]{zhongxiyu@yau.edu.cn}
\affiliation{College of Physics and Electronic Information Engineering, Jining Normal University, Wulanchabu, 012000, China}

\author{Hong-Da Lyu}
\email[]{hongdalyu@sdu.edu.cn}
\affiliation{Key Laboratory of Particle Physics and Particle Irradiation (MOE),
Institute of Frontier and Interdisciplinary Science,
Shandong University, Qingdao, Shandong, 266237, China}

\author{Shoulong Li}
\email[Corresponding author: ]{shoulongli@hunnu.edu.cn}
\affiliation{Department of Physics, Key Laboratory of Low Dimensional Quantum Structures and Quantum Control of Ministry of Education, and Institute of Interdisciplinary Studies, Hunan Normal University, Changsha, 410081, China}
\affiliation{Hunan Research Center of the Basic Discipline for Quantum Effects and Quantum Technologies, Hunan Normal University, Changsha 410081, China}

\author{Hongwei Yu}
\email[]{hwyu@hunnu.edu.cn}
\affiliation{Department of Physics, Key Laboratory of Low Dimensional Quantum Structures and Quantum Control of Ministry of Education, and Institute of Interdisciplinary Studies, Hunan Normal University, Changsha, 410081, China}
\affiliation{Hunan Research Center of the Basic Discipline for Quantum Effects and Quantum Technologies, Hunan Normal University, Changsha 410081, China}


\date{\today}

\begin{abstract}

Hellings-Nordtvedt theory is a vector-tensor theory in which a vector field $A_\mu$ is nonminimally coupled to curvature through  two independent interactions $A^2{\cal R}$ and $A^\mu A^\nu{\cal R}_{\mu\nu}$. When supplemented by a potential whose zero-energy minimum occurs at nonzero $A^2$, the restricted $A^\mu A^\nu{\cal R}_{\mu\nu}$ sector is known to admit black-hole and neutron-star solutions with a {global} monopole-like asymptotic vacuum structure. We examine whether this structure is a generic consequence of the nonzero vector vacuum or instead relies on the special Ricci-tensor coupling. By analyzing the field equations near spatial infinity, we show that the asymptotic vacuum condition is incompatible with generic nonzero values of both couplings and instead selects two allowed single-coupling sectors. The $A^\mu A^\nu{\cal R}_{\mu\nu}$ sector reproduces the known {global}  monopole-like asymptotics, whereas the $A^2{\cal R}$ sector admits an asymptotically flat Schwarzschild metric with a nontrivial radial vector field. We further compute the Noether mass in the $A^2{\cal R}$ sector, derive the corresponding Solar-System constraints, and construct neutron-star configurations. Although the weak-field deviation is constrained to be small, neutron stars can still show appreciable departures from both general relativity and the Ricci-tensor-coupling sector in their masses, radii, and moments of inertia. Our results identify that the $A^2{\cal R}$ sector of massive Hellings-Nordtvedt theory as a viable and useful framework for studying strong-field compact objects with a nonzero vector vacuum while remaining compatible with weak-field tests.
  
\end{abstract}


\maketitle

\section{Introduction}

Nonminimal couplings between spacetime curvature and additional gravitational degrees of freedom provide a systematic way to parametrize possible deviations from general relativity (GR) and to explore gravitational phenomena in regimes where the theory has not yet been directly tested with high precision~\cite{Will:2014kxa,Berti:2015itd,Belenchia:2016bvb,Shankaranarayanan:2022wbx,Heisenberg:2018vsk,Olmo:2019flu,Doneva:2022ewd}. When the additional degree of freedom is a vector field, the resulting theories belong to the broad class of vector-tensor gravity. An early and representative example is the Hellings-Nordtvedt theory~\cite{Hellings:1973zz}, in which a massless vector field $A_\mu$ is coupled nonminimally to curvature through two independent interactions, $A_\mu A^\mu {\cal R}$ and $A^\mu A^\nu {\cal R}_{\mu\nu}$, where ${\cal R}$ and ${\cal R}_{\mu\nu}$ are the Ricci scalar and the Ricci tensor, respectively. These two terms provide a useful setting in which the physical roles of different curvature-vector couplings can be isolated. They have been studied separately in strong-field systems, including black holes~\cite{Xu:2022frb,Mai:2023ggs,Liang:2022gdk,Xu:2023xqh,Mai:2024lgk,Xu:2026zgd} and neutron stars~\cite{Annulli:2019fzq,Ji:2024aeg,Hu:2023vsg,Silva:2021jya,Demirboga:2021nrc}, as well as in related cosmological settings~\cite{Hell:2024xbv,DeFelice:2025ykh,DeFelice:2025khe,Hell:2026sxt}. For a special choice of the coupling constants, the two elementary interactions combine into the Einstein-tensor coupling $A^\mu A^\nu G_{\mu\nu}$, which also appears as a building block in more general vector-tensor theories~\cite{Heisenberg:2014rta,Tasinato:2014eka,Geng:2015kvs,DeFelice:2016yws,Allys:2016jaq,Kimura:2016rzw}.

A natural extension of Hellings-Nordtvedt theory is to supplement the vector sector with a potential. As is familiar from massive scalar-tensor theories~\cite{Alsing:2011er}, giving an additional gravitational degree of freedom a mass can modify its effective range and hence change the interpretation of weak-field and binary-pulsar constraints. In the present vector-tensor context, the potential can play a more special role. If the potential has a vanishing minimum at a nonzero value of the vector invariant $A^2$, the vacuum condition requires the vector field to approach a nonzero value asymptotically. The action remains generally covariant, while the vacuum configuration selects a preferred spacetime direction that persists even in the asymptotic region. This is commonly interpreted as spontaneous Lorentz symmetry breaking~\cite{Kostelecky:2003fs}. Massive Hellings-Nordtvedt theory with such a potential is therefore closely related to bumblebee models of vector-tensor gravity. 

Recently, a simple exact black-hole solution~\cite{Casana:2017jkc} and self-consistent neutron-star solutions~\cite{Luo:2026oxw} were obtained in a restricted sector of the massive Hellings-Nordtvedt theory with such a bumblebee-type potential, where only the coupling $A^\mu A^\nu {\cal R}_{\mu\nu}$ is retained. These solutions share a nontrivial asymptotic structure: the spacetime is not asymptotically flat in the standard sense, but instead contains a solid-angle deficit analogous to that of a global monopole~\cite{Barriola:1989hx}. The deficit angle is described by a so-called Lorentz-violating parameter combining the nonminimal coupling constant with the asymptotic vector amplitude. Although this parameter is stringently constrained by Solar-System tests~\cite{Casana:2017jkc}, and hence the corresponding vector-tensor black hole is expected to differ only slightly from its GR counterpart, the vector sector can still generate appreciable departures from GR in neutron-star structure and associated observables~\cite{Luo:2026oxw}. 

However, it remains unclear whether the global monopole-like asymptotic structure is a consequence of the special potential itself, or whether it relies on the special curvature-vector coupling $A^\mu A^\nu {\cal R}_{\mu\nu}$ adopted in the restricted theory. This distinction is important. If the nonzero vector vacuum selected by the potential were sufficient to determine the asymptotic geometry, one would expect similar global monopole-like behavior to persist in the full Hellings-Nordtvedt theory and, in particular, in the Einstein-tensor sector $A^\mu A^\nu G_{\mu\nu}$. If, on the other hand, the asymptotic structure depends on the specific nonminimal coupling, then one must clarify the role played by the $A^2{\cal R}$ coupling and determine what types of compact-object solutions it can support.

In this work, we address these issues by studying black holes and neutron stars in the full Hellings-Nordtvedt theory with a bumblebee-type potential. We show that the asymptotic vacuum condition selected by the potential does not allow generic nonzero values of the two nonminimal couplings. Instead, the field equations separate the theory into two allowed single-coupling sectors. The $A^\mu A^\nu{\cal R}_{\mu\nu}$ sector reproduces the previously known global monopole-like asymptotic structure, whereas the $A^2{\cal R}$ sector admits asymptotically flat compact-object solutions. Thus, the nonzero vector vacuum, or equivalently the spontaneous breaking of Lorentz symmetry, does not by itself determine the asymptotic geometry; the specific curvature-vector coupling is essential. We further analyze the asymptotically flat $A^2{\cal R}$ sector, focusing on its weak-field constraints and neutron-star configurations, and show that appreciable strong-field deviations may still arise in neutron stars.

The remainder of this manuscript is organized as follows. 
In Sec.~\ref{MHNT}, we review massive Hellings-Nordtvedt theory and derive the field equations. 
In Sec.~\ref{asymptotic}, we analyze the asymptotic vacuum structure and show that a nonzero vector vacuum at spatial infinity, together with the field equations, restricts the theory to the allowed single-coupling sectors.
In Sec.~\ref{noether_solar}, we compute the Noether mass of the black hole in the $A^2{\cal R}$ sector and use the physical mass to derive Solar-System constraints on the corresponding Lorentz-violating parameter. 
In Sec.~\ref{neutronstar}, we construct slowly rotating neutron-star solutions in the $A^2{\cal R}$ sector, analyze their equilibrium properties and moments of inertia, and compare these observables with the corresponding results in the $A^\mu A^\nu{\cal R}_{\mu\nu}$ sector. 
Finally, Sec.~\ref{conclusion} summarizes our results and discusses possible extensions.

\section{Massive Hellings-Nordtvedt theory} \label{MHNT}

We consider the Hellings-Nordtvedt theory of gravity supplemented by a potential for the vector field. The action is given by~\cite{Hellings:1973zz,Kostelecky:2003fs}
\begin{equation}
S = \frac{c^4}{16\pi G}\int d^4x \sqrt{-g} L +S_{\rm m} \,,
\end{equation}
where the Lagrangian density is
\be
L = {\cal R} 
+ \gamma_1 X {\cal R}
+ \gamma_2 A^\mu A^\nu {\cal R}_{\mu\nu} 
-\frac{1}{4}F^2
- V(X) \,.
\label{HN}
\ee
Here $g$ denotes the determinant of the spacetime metric $g_{\mu\nu}$, while ${\cal R}$ and ${\cal R}_{\mu\nu}$ are the Ricci scalar and Ricci tensor, respectively. The vector field is denoted by $A_\mu$, with field strength $F_{\mu\nu}=2\nabla_{[\mu}A_{\nu]}$, and  $X \equiv A^2$.
The function $V(X)$ is the potential of the vector field. {The constants $\gamma_1$ and $\gamma_2$ are dimensionless coupling constants in the units used in this work, and characterize the two nonminimal curvature-vector couplings, respectively.} The matter fields are assumed to be minimally coupled to $g_{\mu\nu}$ through the matter action $S_{\rm m}$. Throughout this paper, we adopt geometric units $G=c=1$.
Variation of the action with respect to $g_{\mu\nu}$ and $A_\mu$ yields the gravitational field equation and the vector field equation, which are given by
\begin{widetext}
\bea
E_{\mu\nu}
&\equiv&
G_{\mu\nu}
+\gamma_1 \left[ X G_{\mu\nu}
+{\cal R} A_\mu A_\nu
+\left(g_{\mu\nu}\Box-\nabla_\mu\nabla_\nu\right)X \right]
+\gamma_2 \Big[2{\cal R}^{\rho}{}_{(\mu}A_{\nu)}A_\rho
\nonumber\\
&\quad& 
-\frac{1}{2}g_{\mu\nu}{\cal R}_{\rho\sigma}A^\rho A^\sigma
+\frac{1}{2}g_{\mu\nu}\nabla_\rho\nabla_\sigma
\left(A^\rho A^\sigma\right)
+\frac{1}{2}\Box\left(A_\mu A_\nu\right)
-\nabla_\rho\nabla_{(\mu}
\left(A_{\nu)}A^\rho\right) \Big]
\nonumber\\
&\quad&
-\frac{1}{2}
\left[
F_{\mu\lambda}F_{\nu}{}^{\lambda}
-\frac{1}{4}g_{\mu\nu}F^2
\right] 
+\frac{1}{2}g_{\mu\nu}V
-V_X A_\mu A_\nu
=8\pi T_{\mu\nu} \,,
\label{eom1}
\\
E_A{}^\nu
&\equiv&
\frac{1}{2}\nabla_\mu F^{\mu\nu}
+\gamma_1 {\cal R} A^\nu
+\gamma_2 {\cal R}^{\mu\nu}A_\mu
-V_X A^\nu
=0 \,.
\label{eom2}
\eea
\end{widetext}
where $G_{\mu\nu} \equiv  {\cal R}_{\mu\nu} - {\cal R} g_{\mu\nu}/2$ is the Einstein tensor, $V_X = dV/dX$, $T_{\mu\nu} = -\frac{2}{\sqrt{-g}} \frac{\delta S_\textup{m}}{\delta g^{\mu\nu}} $ is the energy-momentum tensor, $\nabla_\mu$ is the covariant derivative, and $\Box = \nabla_\mu \nabla^\mu$ is the d'Alembert operator. Parentheses $(\mu\nu)$ and  square brackets $[\mu\nu]$ indicate symmetrization and antisymmetrization over the enclosed indices.

\section{Asymptotic vacuum structure and black holes} \label{asymptotic}

In this section, we first analyze the asymptotic vacuum structure of the full massive Hellings-Nordtvedt theory~\eqref{HN} and show how the special potential restricts the nonminimal couplings. The most general metric ansatz for static and spherically symmetric self-gravitating objects can be written as
\be
ds^2 = - h(r) dt^2 + \frac{dr^2}{f(r)}  + r^2 (d\theta^2 + \sin^2\theta d\varphi^2)  \,,  \label{metric}  
\ee
and the vector field is chosen to have a nonvanishing radial component,
\be
{A=A_r dr=b\phi(r)dr} \,,\quad \textup{with}\quad X = b^2 \phi^2 f \,, \label{vectoransatz}
\ee
where $b$ is a nonzero {dimensionless} constant. Substituting Eqs.~\eqref{metric}--\eqref{vectoransatz} into the field equations~\eqref{eom1}--\eqref{eom2} with $T_{\mu\nu}=0$, we obtain the following independent nonvanishing components of the vacuum equations:
\begin{widetext}
\begin{subequations} \label{vaceq}
\bea
E^t{}_t &\equiv& 
\frac{b^2 f\left[-\gamma_1
+f\left(\gamma_1+\gamma_2\right)\right]\phi^2}{r^2}
+\frac{1}{4}b^2
\left(2\gamma_1+\gamma_2\right)
\phi^2 f'^2
+\frac{4b^2 f^2
\left(\gamma_1+\gamma_2\right)\phi\phi'}{r}
\nonumber\\
&&
+b^2 f^2
\left(2\gamma_1+\gamma_2\right)\phi'^2 +\frac{1}{2}b^2 f
\left(2\gamma_1+\gamma_2\right)\phi^2 f''
+b^2 f^2
\left(2\gamma_1+\gamma_2\right)\phi\phi''
\nonumber\\
&&
+f'
\left[
\frac{3b^2 f
\left(\gamma_1+\gamma_2\right)\phi^2}{r}
+\frac{5}{2}b^2 f
\left(2\gamma_1+\gamma_2\right)\phi\phi'
\right] + \frac{f'}{r} +\frac{f-1}{r^2} +\frac{V}{2}
 = 0 \,, \label{ett} \\
E^r{}_r &\equiv& 
\frac{b^2 f\phi^2}{4r^2h^2}
\bigg\{r^2 f
\left(2\gamma_1+\gamma_2\right)h'^2
-4h^2
\left[
V_X\,r^2-\gamma_1
+f\left(\gamma_1-\gamma_2\right)
\right]
\nonumber\\
&&
-2rfh
\left[
2\left(\gamma_1-\gamma_2\right)h'
+r\left(2\gamma_1+\gamma_2\right)h''
\right]
\bigg\} 
+\frac{
b^2 f^2
\left(2\gamma_1+\gamma_2\right)
\phi\left(4h+r h'\right)\phi'
}{2rh}
\nonumber\\
&&
+\frac{(-1+f)h+r f h'}{r^2 h} +\frac{V}{2}
 = 0 \,, \label{err} \\
E^\theta{}_\theta &=& E^\varphi{}_\varphi \equiv 
\frac{1}{4}b^2
\left(2\gamma_1+\gamma_2\right)
\phi^2 f'^2
+b^2 f^2
\left(2\gamma_1+\gamma_2\right)\phi'^2
+b^2 f^2
\left(2\gamma_1+\gamma_2\right)\phi\phi''
\nonumber\\
&&
+\frac{V}{2} -\frac{
b^2 f^2\left(\gamma_1+\gamma_2\right)
\phi^2 h'\left(-2h+r h'\right)
}{4rh^2}
+\frac{
b^2 f^2\left(\gamma_1+\gamma_2\right)
\phi\left(2h+r h'\right)\phi'
}{rh}
\nonumber\\
&&
+f'
\bigg[
\frac{
3b^2 f\left(\gamma_1+\gamma_2\right)
\phi^2\left(2h+r h'\right)
}{4rh}
+\frac{1}{4}
\left(
\frac{2}{r}+\frac{h'}{h}
\right)
+\frac{5}{2}b^2 f
\left(2\gamma_1+\gamma_2\right)\phi\phi'
\bigg]
\nonumber\\
&&
+\frac{1}{2}b^2 f
\left(2\gamma_1+\gamma_2\right)\phi^2 f''
+\left[
\frac{f}{2h}
+\frac{
b^2 f^2\left(\gamma_1+\gamma_2\right)\phi^2
}{2h}
\right]h''
  -\frac{f h'\left(r h' -2h \right)}{4rh^2} = 0 \,, \label{ethetatheta} \\
{E_A{}^r} &\equiv& 
\frac{b f\phi}{2r^2h^2}
\bigg[
-8(-1+f)h^2\gamma_1
-8rfh\gamma_1 h'
+r^2f\left(2\gamma_1+\gamma_2\right)h'^2
\bigg] -2b\,V_X f\phi
\nonumber\\
&&
-\frac{
b f^2\left(2\gamma_1+\gamma_2\right)\phi h''
}{h} -\frac{
b f\left(2\gamma_1+\gamma_2\right)
\phi f'\left(4h+r h'\right)
}{2rh}
 = 0 \,. \label{eAr} 
\eea
\end{subequations}
\end{widetext}
Here a prime denotes differentiation with respect to $r$.
Instead of solving these equations directly, we first examine them near spatial infinity. We therefore expand the metric functions and the vector-field profile as
\bea
\lim_{r \rightarrow \infty} h(r) &=& \sum_{i=0}^{\infty} \hat{h}_i r^{-i}  \,, \nn \\ 
\lim_{r \rightarrow \infty} f (r) &=& \sum_{i=0}^{\infty} \hat{f}_i r^{-i} \,, \nn\\
\lim_{r \rightarrow \infty} \phi(r) &=& \sum_{i=0}^{\infty} \hat{\phi}_i r^{-i} \,. \label{inftycond}
\eea
Substituting these expansions into Eq.~\eqref{vaceq} and solving the resulting equations order by order in powers of $1/r$, we find that the leading-order vacuum equations are
\bea
E^t{}_t &\equiv& \frac12 V(\hat{X}_0) + {\cal O} (r^{-1}) \,,\nn \\
E^r{}_r &\equiv& -b^2 \hat{\phi}_0^2 \hat{f}_0 V_X(\hat{X}_0) + \frac12 V(\hat{X}_0)  + {\cal O} (r^{-1}) \,,\nn \\
E^\theta{}_\theta &\equiv& \frac12 V(\hat{X}_0) + {\cal O} (r^{-1}) \,,\nn \\
E_A{}^r &\equiv& -2 b \hat{\phi}_0 \hat{f}_0 V_X(\hat{X}_0) + {\cal O} (r^{-1})\,,
\eea
where $\hat{X}_0 = b^2  \hat{\phi}_0^2 \hat{f}_0$. Thus, before any explicit solution is constructed, the leading-order field equations already require both the potential and its first derivative to vanish in the asymptotic vacuum region,
\be
V(\hat{X}_0) = 0 \,, \quad V_X(\hat{X}_0) = 0 \,. \label{potentialcond1}
\ee
At this stage, no explicit functional form of $V(X)$ is needed. One may note that the massless case $V=0$, or a potential whose relevant vacuum occurs at $X=0$, also satisfies the conditions in Eq.~(\ref{potentialcond1}). {However, such choices do not implement the spontaneous Lorentz-symmetry breaking considered in this work, since potential itself does not select a nonzero vector configuration at spatial infinity. We therefore focus on potentials for which the zero-energy minimum is located at a nonzero value $b^2$ of the vector invariant $X$. The asymptotic vacuum condition is then chosen as }
\be
{X}|_{r\to\infty}  =b^2 \,, \quad \textup{with}\quad V|_{X= b^2} = 0 \,, \quad V_X |_{X= b^2} = 0 \,. \label{potentialcond2}
\ee
For the ansatz~\eqref{vectoransatz}, imposing the above condition in the asymptotic expansion gives
\be
\lim_{r \rightarrow \infty} \phi(r)=\lim_{r \rightarrow \infty} f (r)^{-1/2} \,.
\label{phifcond}
\ee
{Under this branch condition, the vacuum field equations~\eqref{vaceq} can be solved directly, without performing an additional order-by-order expansion based on Eq.~\eqref{inftycond}. However, since the branch condition~\eqref{phifcond} itself is obtained from the large-$r$ asymptotic analysis, the resulting configurations should be understood as asymptotic branches. Their asymptotic consistency selects two single-coupling sectors.}
The corresponding asymptotic configurations are
\bea
\textup{Case I}\ (A^2{\cal R}):  &&  \quad  \gamma_2 = 0 \,,\quad \ell_1 = \gamma_1 b^2 \,,
 \nn \\ 
&&\quad 
 h(r) = 1 - \frac{m}{r }  \,, \nn \\ 
&&\quad 
 f(r)  =  1 - \frac{m}{r }  \,, \nn \\ 
&&\quad 
\phi(r)  =   \left(1 - \frac{m}{r } \right)^{-\frac12} \,, \label{inftycond1} \\
\textup{Case II}\ (A^\mu A^\nu{\cal R}_{\mu\nu}): &&\quad \gamma_1 = 0 \,,\quad \ell_2 = \gamma_2 b^2 \,,\nn \\ 
&&\quad 
 h(r) = 1 - \frac{m}{r }  \,,  \nn \\ 
&&\quad 
 f(r)  = \frac{1}{1 + \ell_2} \left[1 - \frac{m}{r } \right] \,, \nn \\ 
&&\quad 
 \phi(r)  =  \left[\frac{1 + \ell_2}{1 - \frac{m}{r }}\right]^{\frac12}  \,. \label{inftycond2} 
\eea
Here $m$ is an integration constant, while $\ell_1$ and $\ell_2$ are the corresponding Lorentz-violating parameters. These results show that the asymptotic vacuum branch selected by the potential, together with the field equations, separates the two nonminimal curvature-vector couplings into two distinct allowed sectors. In particular, the two couplings $\gamma_1$ and $\gamma_2$ cannot be nonzero simultaneously on this branch. The commonly used Einstein-tensor coupling, being a nontrivial linear combination of the two curvature-vector couplings, is therefore excluded by the same asymptotic consistency requirement. The two allowed single-coupling sectors have qualitatively different asymptotic geometries. Case II, in which only the coupling $A^\mu A^\nu{\cal R}_{\mu\nu}$ is present, reproduces the global monopole-like asymptotic structure studied in previous work~\cite{Casana:2017jkc}. By contrast, Case I, in which only the coupling $A^2{\cal R}$ is present, also admits vacuum solutions but remains asymptotically flat. Therefore, the global monopole-like asymptotics is not generated by the potential alone; rather, it depends crucially on the specific form of the nonminimal curvature-vector coupling.

If the branch condition $X=b^2$ is imposed throughout the exterior vacuum region, the asymptotic configurations in Eqs.~\eqref{inftycond1} and~\eqref{inftycond2} solve the full field equations and become exact black-hole solutions. In Case I, where only the $A^2{\cal R}$ coupling is present, the metric is identical to the Schwarzschild solution, whereas the vector field remains nontrivial. Black holes with this property are usually referred to as stealth solutions in vector-tensor theories~\cite{Chagoya:2016aar,Minamitsuji:2016ydr,Chagoya:2017ojn,Xu:2026zgd,Bakopoulos:2024zke}, and analogous examples have also been found in theories containing other nonminimally coupled fields~\cite{Babichev:2013cya,Minamitsuji:2023nvh,Bakopoulos:2024zke,Yu:2025odj} and higher-curvature pure gravity~\cite{Lu:2015cqa,Liu:2020yqa,Li:2017ncu,Li:2023vbo,LiangLiang:2026bgx}. However, thermodynamic analyses of stealth black holes in some theories have shown that the relation between the integration constant appearing in the metric and the conserved mass can be modified by the nontrivial additional field~\cite{Bakopoulos:2024zke,Yu:2025odj}. Therefore, the physical mass should be treated with care and determined using a more general method, rather than inferred directly from the GR mass definition. This distinction is important because the physical mass, rather than the metric integration constant, should be used when deriving observational constraints on the theoretical parameters from the solution~\cite{Yu:2025odj}.

\section{Noether Mass and Solar-System Constraints} \label{noether_solar}

In this section, we focus on the stealth black hole in Case I~(\ref{inftycond1}). We first evaluate its Noether mass and show that the Lorentz-violating parameter $\ell_1$ gives a nontrivial correction to the physical mass, thereby breaking the apparent degeneracy with the Schwarzschild black hole at the level of conserved charges. We then use this physical mass to derive Solar-System constraints on the Lorentz-violating parameter $\ell_1$.

\subsection{Noether mass} \label{noethermass}

For this purpose, we use the well-known Wald covariant phase-space formalism~\cite{Wald:1993nt,Iyer:1994ys}, which has been widely applied to gravitational theories coupled to vector fields~\cite{Fan:2017bka,Gao:2003ys,An:2024fzf,Chen:2025ypx,Wu:2015ska,Li:2025tcd,Li:2016nll}. We only recall the ingredients needed for the computation of the Noether mass and then apply them to the Case I black hole~(\ref{inftycond1}).

For a Lagrangian $D$-form $\mathbf{L}(\psi)$, its variation takes the form
\be
\delta \mathbf{L} = \mathbf{E}_\psi \delta \psi + d \mathbf{\Theta} (\psi, \delta \psi) \,, \nn
\ee
where $\mathbf{\Theta}$ is the symplectic potential. The Noether current associated with  $\xi$ is defined by
\begin{equation}
\mathbf{J}_\xi = \mathbf{\Theta}(\psi,\mathscr{L}_\xi \psi) - i_\xi \mathbf{L}\,,\nn
\end{equation}
and satisfies $\mathbf{J}_\xi = d\mathbf{Q}_\xi$ on shell. The variation of the Hamiltonian charge associated with $\xi$ is then given by
\begin{equation}
\delta \mathcal{H} = \frac{1}{16\pi} \int_{\Omega_{D-2}} \left( \delta \mathbf{Q}_\xi - i_\xi \mathbf{\Theta} \right)\,.\nn
\label{WaldH}
\end{equation}
The first law follows by equating the Hamiltonian variation evaluated at spatial infinity and at the horizon, namely $\delta \mathcal{H}_\infty = \delta \mathcal{H}_{r_h}$.

For the full massive Hellings-Nordtvedt theory~\eqref{HN} in four dimensions, a direct computation gives
\begin{equation}
\delta \mathbf{Q}_\xi - i_\xi \mathbf{\Theta} 
= -\frac{4\pi r\sqrt{h}}{\sqrt{f}} (1+\ell_1+\ell_2)\, \sin\theta \, \delta f \,.
\end{equation}
For the Case I~(\ref{inftycond1}), the Noether mass is
\be
M_1 = \frac12 (1+\ell_1)m \,.
\label{NoetherM1}
\ee
For comparison,we also give the Noether mass in Case II~\eqref{inftycond2}, 
\be
M_2 = \frac{ \sqrt{1+\ell_2}m}{2} \,, \nn
\label{NoetherM2}
\ee
which reproduces the result obtained in the Ricci-tensor-coupling sector~\cite{An:2024fzf,Chen:2025ypx,Li:2025tcd}. Therefore, in both allowed sectors, the physical mass differs from the geometric parameter $m/2$ by a coupling-dependent factor. 
In terms of the physical mass $M_1$, the Case I black-hole solution in the $A^2{\cal R}$ sector can be rewritten as
\begin{widetext}
\bea
ds^2 &=& -\left[1 - \frac{2 M_1}{r (1+\ell_1)}\right] dt^2 + \left[1 - \frac{2 M_1}{r (1+\ell_1)} \right]^{-1}dr^2 + r^2 (d\theta^2 +\sin^2\theta d\varphi^2) \,,\nn \\
A &=& b \left[1 - \frac{2 M_1}{r (1+\ell_1)} \right]^{-\frac12} dr \,,\quad \ell_1 = \gamma_1 b^2  \,,\quad \gamma_2 = 0 \,.  \label{BH1a} 
\eea
\end{widetext}
This form makes clear that the apparent degeneracy with the Schwarzschild black hole is broken once the solution is expressed in terms of the physical mass. Since the Lorentz-violating parameter $\ell_1$ is fixed by the product of the nonminimal coupling $\gamma_1$ and the nonzero vector vacuum $b$, this correction is a model-dependent effect of the coupling and potential rather than a universal property of stealth black holes~\cite{Chagoya:2016aar,Minamitsuji:2016ydr,Chagoya:2017ojn,Xu:2026zgd,Bakopoulos:2024zke} in vector-tensor gravity. As a result, the effects of the nonzero vector vacuum and the nonminimal curvature-vector coupling are no longer completely hidden at the level of physical observables.

\subsection{Solar-System Constraints} \label{solar0}

We now derive Solar-System constraints on the Lorentz-violating parameter $\ell_1$ in the sector with only the $A^2{\cal R}$ coupling, with particular emphasis on the role of the Noether mass in the weak-field limit. It is useful to first recall the situation in the sector with only the coupling $A^\mu A^\nu{\cal R}_{\mu\nu}$. Bounds on the corresponding parameter $\ell_2$ were estimated in Ref.~\cite{Casana:2017jkc}. Although the mass parameter used there was not identified through a Noether charge, this ambiguity does not significantly affect the order of magnitude of the resulting weak-field bounds. The reason is that the Ricci-tensor-coupling sector has a global monopole-like asymptotic structure, which already places very stringent constraints on $\ell_2$, typically at the level of $10^{-10}$ or below according to the Solar-System tests considered in Ref.~\cite{Casana:2017jkc}. The correction associated with the precise definition of the conserved mass is therefore subleading in that case. A similar observation was first made in Einstein-Kalb-Ramond gravity~\cite{Yu:2025odj}. 

The situation is qualitatively different in the sector with only the $A^2{\cal R}$ coupling. In this case, the black-hole metric is exactly Schwarzschild when written in terms of the integration constant, and the Lorentz-violating parameter enters the weak-field geometry only after the solution is expressed in terms of the physical mass. Thus, the Noether mass is not merely a refinement of the weak-field analysis; it is essential for extracting any metric-based constraint on $\ell_1$. From Eq.~\eqref{BH1a}, the effective mass entering the weak-field observables is
\be
M_{\rm eff}=\frac{M_1}{1+\ell_1},
\label{Meff}
\ee
The derivation of the Solar-System constraints then reduces to comparing the standard GR predictions written in terms of $M_{\rm eff}$ with those written in terms of the physical mass $M_1$.

Following the same three standard Solar-System tests considered in Refs.~\cite{Casana:2017jkc,Yang:2023wtu}, namely the perihelion advance of Mercury, the deflection of light, and the Shapiro time delay, we derive the allowed intervals for $\ell_1$ in massive Hellings-Nordtvedt theory with only the $A^2{\cal R}$ coupling. In contrast to Ref.~\cite{Casana:2017jkc}, where the emphasis was mainly on upper bounds, we follow the approach of Ref.~\cite{Yang:2023wtu} by using representative observational results to infer the corresponding allowed ranges of $\ell_1$. Throughout this analysis, the static and spherically symmetric metric is parametrized by the Noether mass. These intervals will be used in the next section to choose representative values of $\ell_1$ for studying strong-field deviations in neutron-star configurations.

We first consider the perihelion advance of Mercury. The anomalous advance is usually expressed as the accumulated angular shift per Julian century. In GR, the correction relative to Newtonian gravity is denoted by $\Delta\Phi_{\rm GR}$. For the present $A^2{\cal R}$ coupling, the weak-field metric differs from the GR Schwarzschild metric only through the mass replacement in Eq.~\eqref{Meff}. In the parametrization used here, the perihelion correction scales as the square of the effective mass parameter. The predicted perihelion advance in massive Hellings-Nordtvedt theory is therefore
\be
\Delta \Phi_{\rm HN1} = \frac{\Delta \Phi_{\rm GR}}{(1+\ell_1)^2} \,.
\ee
For Mercury, the GR prediction is $\Delta \Phi_{\rm GR}=42.9814''$, while the observed value is $\Delta \Phi_{\rm obs}=(42.9794\pm0.0030)''$~\cite{Casana:2017jkc}, with both values given per Julian century. The perihelion advance of Mercury then gives
\be
-1.2 \times 10^{-5}  \le \ell_1 \le 5.8 \times 10^{-5} \,.  \label{constraint1}
\ee

Next, we consider the light-deflection test. The observable is the deflection angle of a light ray grazing the solar surface. In GR, this angle is denoted by $\Delta\Psi_{\rm GR}$ and is proportional to the mass parameter in the leading weak-field expression. Therefore, in massive Hellings-Nordtvedt theory with only the $A^2{\cal R}$ coupling, the predicted deflection angle is
\be
\Delta \Psi_{\rm HN1} = \frac{\Delta \Psi_{\rm GR}}{1+\ell_1} \,.
\ee
For this test, we use the observational accuracy inferred from very-long-baseline interferometry measurements for a ray grazing the Sun. The GR prediction is $\Delta\Psi_{\rm GR}=1.7516687''$, while the observed light deflection at the solar surface is written as $(1+\gamma_{\rm PPN})\Delta\Psi_{\rm GR}/2$, with $\gamma_{\rm PPN}=0.99992\pm0.00012$~\cite{LambertPoncin}. This gives
\be
-2 \times 10^{-5} \le \ell_1 \le  1.0 \times 10^{-4} \,. \label{constraint2}
\ee

Finally, we consider the Shapiro time-delay test. In GR, the excess propagation time of a radar signal passing near the Sun is denoted by $\Delta T_{\rm GR}$ and is proportional to the mass parameter in the leading weak-field expression. For the present $A^2{\cal R}$ coupling, the predicted time delay is therefore
\be
\Delta T_{\rm HN1} = \frac{\Delta T_{\rm GR}}{1+\ell_1} \,.
\ee
We use the Cassini measurement of the post-Newtonian parameter $\gamma_{\rm PPN}$, for which the observed time delay is written as $(1+\gamma_{\rm PPN})\Delta T_{\rm GR}/2$, with $\gamma_{\rm PPN}=1+(2.1\pm2.3)\times10^{-5}$~\cite{Bertotti:2003rm}. The resulting constraint is
\be
-2.2 \times 10^{-5}  \le \ell_1 \le 1.0 \times 10^{-6} \,. \label{constraint3}
\ee

Combining the three Solar-System constraints in Eqs.~\eqref{constraint1}, \eqref{constraint2}, and \eqref{constraint3}, we find that the Lorentz-violating parameter $\ell_1$ in massive Hellings-Nordtvedt theory with only the $A^2{\cal R}$ coupling is constrained roughly within the range $-10^{-5}\lesssim \ell_1 \lesssim 10^{-6}$. In this sector, the Noether charge does not merely provide a quantitative correction to the weak-field analysis; rather, it qualitatively determines whether a metric-based constraint on $\ell_1$ can be extracted at all. Although the bound is still stringent, it is several orders of magnitude weaker than the corresponding bound on $\ell_2$ in the sector with only the $A^\mu A^\nu{\cal R}_{\mu\nu}$ coupling. 

It is also worth commenting on a possible interpretation of the Case I black-hole solution~(\ref{BH1a}). By a suitable reparametrization, the metric in Eq.~\eqref{BH1a} can be brought into a form conformally related to the Schwarzschild black hole. This suggests that the $A^2{\cal R}$ coupling may be viewed, at least formally, as admitting an Einstein-frame description in which the gravitational sector is minimally coupled. From the black-hole perspective alone, this may further obscure the distinction from GR. However, such a frame transformation generally transfers the nonminimal coupling from the curvature sector to the matter sector. As a result, compact stars need not remain degenerate with their GR counterparts, because the matter fields inside the star can still couple nonminimally to the vector vacuum. It is therefore still necessary to study whether neutron stars in the $A^2{\cal R}$ sector can exhibit sizable strong-field deviations despite the stringent weak-field bound. In the next section, we take representative values of $\ell_1$ around $10^{-6}$ to address this question.

\section{Neutron stars} \label{neutronstar}

\subsection{Basic equations and boundary conditions} \label{framework}

We now consider neutron-star configurations in the massive Hellings-Nordtvedt theory with only the $A^2{\cal R}$ coupling. For a slowly and uniformly rotating star with angular velocity $\Omega$, the spacetime can be written, to first order in the rotation, as
\be
ds^2 = - e^{\lambda(r)} dt^2 + f(r)^{-1} dr^2  +r^2 (d\theta^2 + \sin^2\theta d\varphi^2) -2\epsilon (\Omega - w(r)) r^2 \sin^2\theta dt d\varphi \,, \label{metricstar}
\ee
where $\epsilon$ is a bookkeeping parameter for the slow-rotation expansion. We use $\lambda$ instead of $h$ for later algebraic convenience. With the present convention, $\Omega-w$ describes the frame-dragging angular velocity, while $w$ is the angular velocity of the fluid relative to the local inertial frame. {In the linear slow-rotation approximation adopted here, we keep only terms up to order ${\cal O}(\Omega)$. Under the assumption that the vector field preserves the spherical symmetry of the nonrotating background, no independent axial vector perturbation is introduced at this order.} Consequently, the rotational corrections to the static metric functions, the vector field, the density, and the pressure enter only at order ${\cal O}(\Omega^2)$ and are therefore neglected here. {At order ${\cal O}(\Omega)$, the only new contribution is the frame-dragging metric function. } The stellar matter is modeled as a perfect fluid,
\be
T^{\mu\nu} = (\rho + p) u^\mu u^\nu + p g^{\mu\nu} \,,
\ee
with four-velocity
\be
u^\mu = (u^t, 0, 0, \epsilon\Omega u^t) \,,  \quad \textup{with}  \quad u^\mu u_\mu = -1 \,.
\ee
The pressure and density are functions of $r$ and are related by the equation of state (EOS)
\be
p = P ( \rho ) \,. \label{eos}
\ee
Substituting Eqs.~(\ref{vectoransatz}), (\ref{metricstar})--(\ref{eos}) into the field equations~(\ref{eom1})--(\ref{eom2}), and expanding order by order in $\epsilon$, the nonvanishing zeroth-order Tolman-Oppenheimer-Volkoff  equations are
\begin{subequations} \label{zeroordertov}
\bea
E^t{}_t &=&   -8\pi \rho \,, \label{etttov} \\
E^r{}_r &=&   8\pi P(\rho) \,, \label{errtov} \\
E^\theta{}_\theta &=&  E^\varphi{}_\varphi =  8\pi P(\rho) \,, \label{ethetathetatov}  \\
E_A{}^r &=&  0 \,. \label{eArtov} 
\eea
\end{subequations}
Here the explicit expressions of $E^\mu{}_\mu$ and $E_A^\nu$ are obtained from Eq.~\eqref{vaceq} by setting $\gamma_1=\ell_1/b^2$ and $\gamma_2=0$.
At first order in $\epsilon$, the only independent equation is $E^t{}_{\varphi}$, which gives
\begin{widetext}
\bea
E^t{}_\varphi &\equiv&  w'' - \frac{f(r \lambda' -3 r \ell_1 \phi^2 f' -8) - r f' +  \ell_1 \phi f^2 (\phi( r \lambda' - 8) -4 r \phi')}{2 r f (1 +  \ell_1 \phi^2 f)} w' \nn \\ 
&\quad& - \frac{16 \pi (\rho + P(\rho))}{ f (1 +  \ell_1 \phi^2 f)} w = 0 \,.  \label{firstorder}
\eea
\end{widetext}
The zeroth-order equations can be arranged as a coupled system of nonlinear ordinary differential equations (ODEs),
\bea
\lambda^{\prime\prime} &=& F_1 (r, \lambda^{\prime}, f, \phi, \rho) \,, \nn\\
f^{\prime} &=& F_2 (r, \lambda^{\prime}, f, \phi, \rho) \,, \nn\\
\phi^{\prime} &=& F_3 (r, \lambda^{\prime}, f, \phi, \rho) \,, \nn\\
\rho^{\prime} &=& F_4 (r, \lambda^{\prime},  \rho) \,.  \label{zeroorder0}
\eea
The first-order equation for the rotational perturbation is a second-order linear homogeneous ODE,
\be
w^{\prime\prime} + F_5(r, \lambda^{\prime}, f, \phi, \rho) w^\prime  + F_6(r, \lambda^{\prime}, f, \phi, \rho) w = 0 \,,  \label{firstorder0}
\ee
The functions $F_i$ denote lengthy expressions that are not displayed explicitly. 

We now specify the boundary conditions for Eqs.~(\ref{zeroordertov})--(\ref{firstorder}). Near the center of the star, the regular solution can be expanded in powers of $r$ as follows:
\bea
\lim_{r \rightarrow 0} \lambda(r) &=& \sum_{i=0}^{\infty} \lambda_i r^i  \,, \nn \\
\lim_{r \rightarrow 0} f (r) &=& \sum_{i=0}^{\infty} f_i r^i \,, \nn \\ 
\lim_{r \rightarrow 0} \phi(r) &=& \sum_{i=0}^{\infty} \phi_i r^i\,, \nn \\
\lim_{r \rightarrow 0} \rho(r) &=& \sum_{i=0}^{\infty} \rho_i r^i\,, \nn \\
\lim_{r \rightarrow 0}  w(r) &=& \sum_{i=0}^{\infty} w_i r^i \,. \label{centercond}
\eea
The central density $\rho_0$ is chosen as an input parameter, while $\lambda_0$, $\phi_0$, and $w_0$ are free parameters fixed by the asymptotic conditions. The leading nontrivial coefficients include
\bea
\lambda_2 &=& \frac{16\pi \ell_1 (2 \rho_0 +3 P(\rho_0)) -V(X_0)}{18 (1 + \ell_1 \phi_0^2)} 
-\frac{b^2 V_X (X_0)}{18 \ell_1 }  \,, \nn \\
f_0 &=& 1  \,, \quad w_2 = \frac{8\pi w_0 (\rho_0 + P(\rho_0))}{5 (1 + \ell_1 \phi_0^2)}  \,, \quad X_0 = b^2 \phi_0^2 \,. \label{centercond1}
\eea
It is worth emphasizing that the difference between the two single-coupling sectors already appears at the level of the central boundary conditions. In the $A^\mu A^\nu{\cal R}_{\mu\nu}$ coupling sector, the solid-angle deficit is not merely an asymptotic property; it is also reflected in the near-center expansion of neutron-star solutions, for which the leading coefficient of the radial metric function satisfies $f_0\ne 1$~\cite{Luo:2026oxw}. In contrast, the $A^2{\cal R}$ coupling sector is asymptotically flat at spatial infinity and obeys the standard central regularity condition $f_0=1$, as in GR.
For a given EOS and central density, the equations~(\ref{zeroorder0})--(\ref{firstorder0}) are integrated from the center to the stellar surface, defined by
\be
p(R) = 0 \,. \label{surfacecond}
\ee
Once the integration reaches the stellar surface at $r=R$, the obtained values of $(\lambda, f, \phi, w)$ are taken as the initial conditions for the exterior integration. These are fixed by the continuity conditions at the stellar surface,
\bea
\lambda_\textup{in}(R) &=& \lambda_\textup{ext}(R) \,, \nn \\
f_\textup{in}(R) &=& f_\textup{ext}(R) \,,  \nn \\
\phi_\textup{in}(R) &=& \phi_\textup{ext}(R) \,,  \nn \\
w_\textup{in}(R) &=& w_\textup{ext}(R) \,,  \label{continue}
\eea
where the subscripts ``in'' and ``ext'' denote the interior and exterior solutions, respectively. The first derivatives of these functions are likewise continuous at $r=R$, although not written explicitly here. Outside the star, the pressure and density vanish. The vacuum equations are obtained from Eqs.~\eqref{zeroordertov}--\eqref{firstorder} by setting $p=\rho=0$, or equivalently from Eq.~\eqref{vaceq} with $\gamma_1=\ell_1/b^2$, $\gamma_2=0$, and $h=e^\lambda$. These equations can be rearranged as
\bea
\lambda^{\prime\prime} &=& \hat{F}_1 (r, \lambda^{\prime}, f, \phi) \,, \nn\\
f^{\prime} &=& \hat{F}_2 (r, \lambda^{\prime}, f, \phi) \,, \nn\\
\phi^{\prime} &=& \hat{F}_3 (r, \lambda^{\prime}, f, \phi) \,,   \label{zeroorder0vac}
\eea
and the first-order rotational equation becomes
\be
w^{\prime\prime} + \hat{F}_4(r, \lambda^{\prime}, f, \phi) w^\prime  = 0 \,.  \label{firstorder0vac}
\ee
The functions $\hat{F}_i$ denote the corresponding vacuum expressions. At spatial infinity, the exterior solution approaches the asymptotically flat black-hole branch obtained in the previous section,
\bea
\lim_{r \rightarrow \infty}  \lambda(r) &=& \log \left[1 - \frac{2 M_1}{r (1 + \ell_1)} \right] \,, \nn \\ 
\lim_{r \rightarrow \infty}  f(r)  &=& 1 - \frac{2 M_1}{r (1 + \ell_1)}  \,, \nn \\ 
\lim_{r \rightarrow \infty}  \phi(r)  &=&  \left[1 - \frac{2 M_1}{r (1 + \ell_1)} \right]^{-\frac12} \,.  \label{inftycond1dd}
\eea
Here $M_1$ is the Noether mass defined in the preceding section. {The above equations already specify the asymptotic behavior of the integration variables, and no further large-$r$ expansion of them is required.}

As discussed above, this asymptotic behavior is compatible with potentials whose zero-energy minimum is located at $X=b^2$. In the black-hole case, the solution belongs to a special vacuum branch for which $X=b^2$ holds throughout the vacuum region. By contrast, for neutron stars the exterior vacuum solution is determined by matching to an interior matter configuration, where the fluid, metric, and vector field interact nontrivially. The exterior boundary data at the stellar surface are therefore different from the black hole spacetime~\cite{Luo:2026oxw,Yu:2025odj,Lessa:2025kln}, and the condition $X=b^2$ should be imposed as an asymptotic vacuum condition rather than as a constraint throughout the exterior region. Therefore, while the preceding weak-field and black-hole analyses only require the potential to have a zero-energy minimum at $X=b^2$ and do not depend on its detailed functional form, the neutron-star interior and exterior solutions generally depend on the explicit choice of the potential. In the numerical analysis below, we adopt a simple realization of this requirement,
\be
V(X) = \alpha (\gamma_1^2+\gamma_2^2) (X - b^2)^2 \,, \label{potentialfunction}
\ee
where $\alpha$ characterizes the mass scale of the vector field. This form can be written in the full massive Hellings-Nordtvedt theory and reduces consistently to the corresponding single-coupling expression in each allowed sector. After rewriting the equations in terms of $\ell_1=\gamma_1 b^2$ and setting $\gamma_2 = 0$, the constant $b$ can be absorbed from the system to be solved.
Substituting the asymptotic background into the ${\cal O} (\epsilon)$ vacuum equation gives
\be
\lim_{r \rightarrow \infty}  w(r) = \Omega - \frac{2 J}{r^3} \,,\label{inftycond2dd}
\ee
where $J$ is the angular momentum of the star. The moment of inertia is therefore defined by
\be
I = \frac{J}{\Omega} \,. \label{moi}
\ee 
For the numerical calculations, we nondimensionalize the variables using the solar mass as the reference scale. The characteristic scalings are
\be
M_\odot  \sim r_\star \sim   p_\star^{-\frac12} \sim \rho_\star^{-\frac12} \sim \alpha_\star^{-\frac12} \sim I_\star^{\frac13} \,. \label{dimension}
\ee
In centimeter-gram-second (cgs) units, the corresponding characteristic quantities are
\bea
r_\star &=& \frac{G M_{\odot}}{c^2} = 1.47 \times 10^{5} \textup{cm} \,, \nn\\
\rho_\star &=& \frac{c^6}{G^3 M_{\odot}^2 } = 6.18\times 10^{17} \textup{g} \cdot \textup{cm}^{-3} \,, \nn\\
p_\star &=& \frac{c^8}{G^3 M_{\odot}^2 } =5.55\times 10^{38}  \textup{g} \cdot \textup{cm}^{-1} \cdot \textup{s}^{-2} \,, \nn\\
I_\star &=& \frac{G^2 M_{\odot}^3}{c^4} = 4.34\times 10^{43}  \textup{g} \cdot \textup{cm}^{2}  \,.
\eea

\subsection{Stellar mass, radius, and moments of inertia}  \label{Results}

Having specified the field equations and the boundary conditions at the center, at the stellar surface, and at spatial infinity in massive Hellings-Nordtvedt theory with only the $A^2{\cal R}$ coupling, we now close the system by choosing an EOS. In this work, we adopt the realistic SLy (Skyrme Lyon) EOS~\cite{Douchin:2001sv,Haensel:2004nu} as a representative example. This choice also allows a direct comparison with the $A^\mu A^\nu{\cal R}_{\mu\nu}$ coupling sector studied in Ref.~\cite{Luo:2026oxw}. The resulting nonlinear ODE system is solved as a boundary-value problem by means of a shooting method, as in Refs.~\cite{Luo:2026oxw,Liu:2024wvw,Li:2025gna,Li:2023vbo,LiangLiang:2026bgx}. Since $\lambda$ enters the equations only through its first derivative, we solve directly for $\lambda'$ rather than for $\lambda$ itself. The shooting parameters are adjusted so that the numerical solution matches the asymptotic conditions for $(\lambda',f,\phi,w,w')$ at a sufficiently large radius. In practice, the outer boundary is chosen large enough that the gravitational mass is accurate to at least two decimal places.

Motivated by the Solar-System constraints derived in the previous section, we take a representative value close to the strongest weak-field upper bound, $\ell_1=10^{-6}$, in the numerical analysis. To make a direct comparison with the $A^\mu A^\nu{\cal R}_{\mu\nu}$ coupling sector studied in Ref.~\cite{Luo:2026oxw}, we use the same choices of the potential parameter $\alpha$, namely $\alpha=10^{-4}\alpha_\star$ and $\alpha=10^{-2}\alpha_\star$.

\begin{figure}[]
\includegraphics[width=0.95\linewidth]{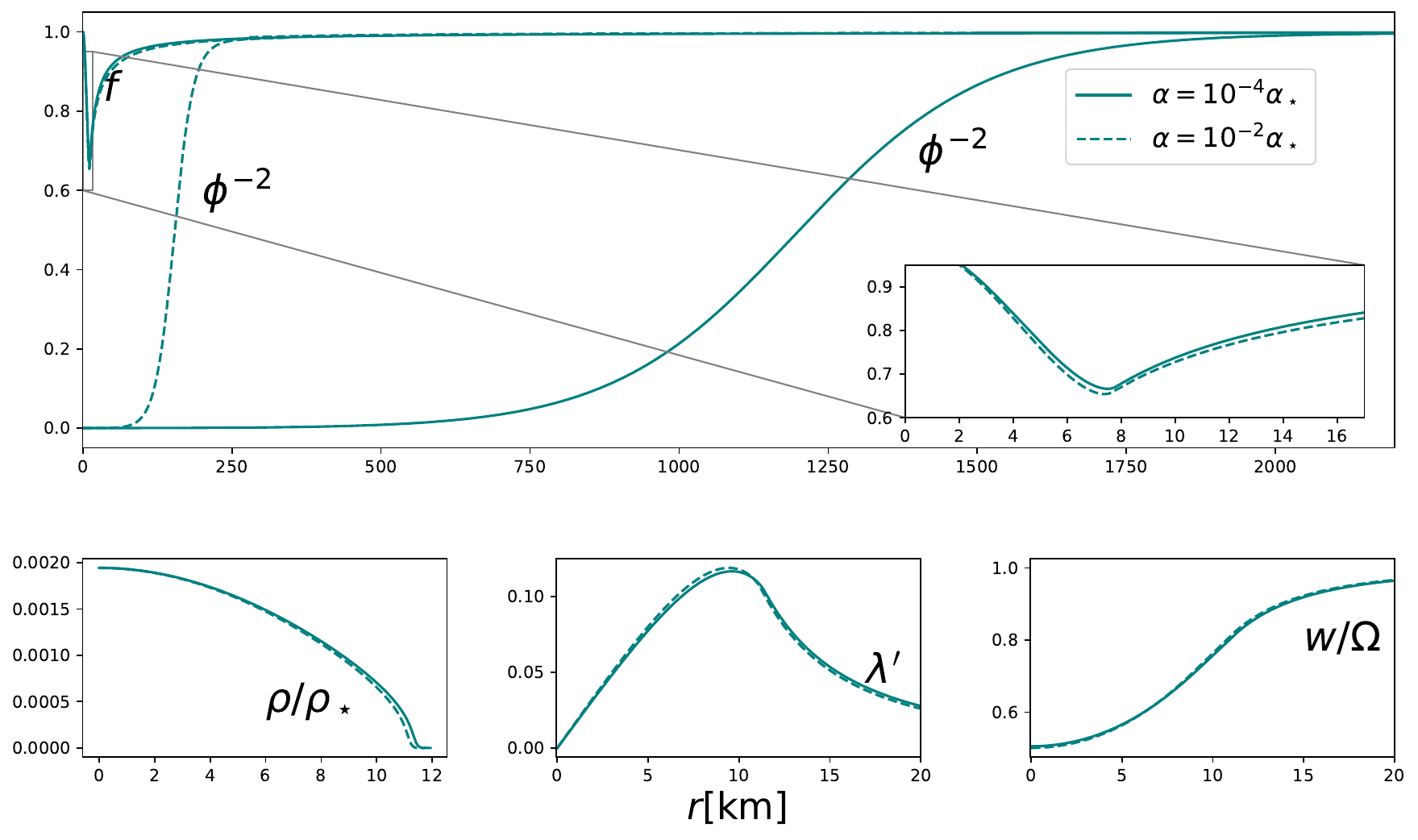}
\caption{\label{fig:0thsolution}The plots illustrate the numerical solutions of the functions 
$(f,\phi^{-2},\rho,\lambda',w/\Omega)$ in massive Hellings-Nordtvedt theory with only the $A^2{\cal R}$ coupling. The solid and dashed lines correspond to $\alpha=10^{-4}\alpha_\star$ and $\alpha=10^{-2}\alpha_\star$, respectively. We fix $\ell_1=10^{-6}$, set the central density to $\rho_0=1.2\times10^{15}\,\textup{g/cm}^3$, and adopt the SLy EOS. The corresponding stellar quantities are $(M,R,I)=(1.71M_\odot,8.12r_\star,47.87I_\star)$ for $\alpha=10^{-4}\alpha_\star$ and $(M,R,I)=(1.66M_\odot,7.98r_\star,43.09I_\star)$ for $\alpha=10^{-2}\alpha_\star$, with the radii corresponding to $11.9\,{\rm km}$ and $11.7\,{\rm km}$, respectively.}
\end{figure}

We first display in Fig.~\ref{fig:0thsolution} representative neutron-star solutions satisfying the required regularity and asymptotic boundary conditions for two values of the potential parameter $\alpha$. 
This serves as a check of the validity of our numerical construction. The upper panel of Fig.~\ref{fig:0thsolution} shows that $f$ and $\phi^{-2}$ coincide only in the weak-field region and asymptotically at spatial infinity, as required by the asymptotic vacuum condition. In the stellar interior and the strong-field region, however, $\phi^2 f$ deviates from unity, and this deviation becomes more pronounced for smaller values of the potential parameter $\alpha$.

\begin{figure}[]
\includegraphics[width=0.95\linewidth]{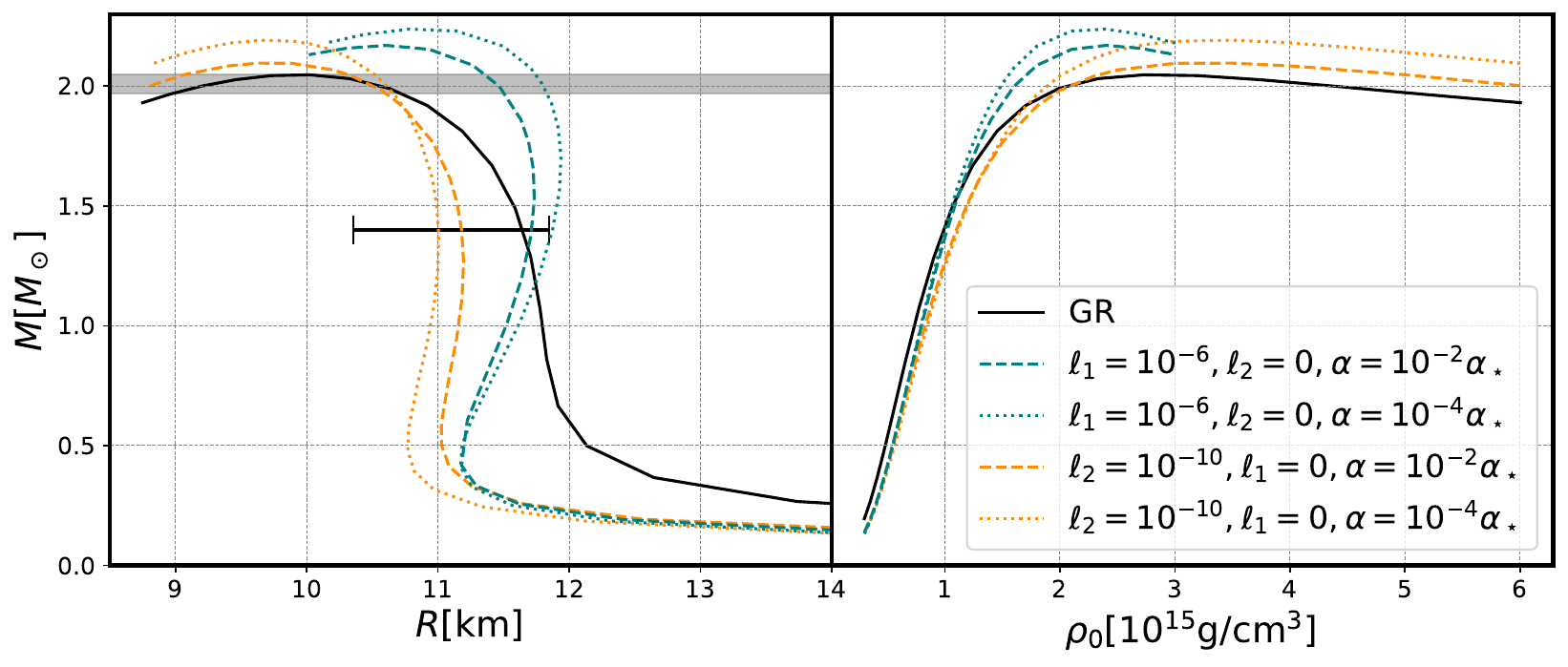}
\caption{\label{fig:MRrelation}Mass-radius ($M$-$R$, left panel) and mass-central density ($M$-$\rho_0$, right panel) relations for neutron stars described by the SLy EOS in massive Hellings-Nordtvedt theory. We compare the two single-coupling sectors: the $A^2{\cal R}$ sector with $(\ell_1,\ell_2)=(10^{-6},0)$ and the $A^\mu A^\nu{\cal R}_{\mu\nu}$ sector with $(\ell_1,\ell_2)=(0,10^{-10})$. The data for the latter sector are taken from Ref.~\cite{Luo:2026oxw}.  The gray shaded region represents the observational mass constraint from a massive neutron star, $M=2.01\pm0.04M_\odot$~\cite{Antoniadis:2013pzd}. The error bar denotes the GW170817 radius constraint for a $1.4M_\odot$ neutron star, $R_{1.4}=11.0^{+0.9}_{-0.6}\,{\rm km}$, at the $90\%$ credible level~\cite{Capano:2019eae}.}
\end{figure}

Next, we examine the mass-radius relations for the SLy EOS by varying the central density $\rho_0$, as shown in Fig.~\ref{fig:MRrelation}. 
For comparison, we also show the corresponding mass-radius curves in the $A^\mu A^\nu{\cal R}_{\mu\nu}$ coupling sector with $\ell_2=10^{-10}$~\cite{Luo:2026oxw}. Here $\ell_1$ and $\ell_2$ are chosen as representative values motivated by their respective weak-field constraints, rather than as equal-strength couplings. Although $\ell_1=10^{-6}$ is already very small, the $A^2{\cal R}$ coupling still produces a noticeable departure from the GR prediction. The qualitative trend is similar to that found in the $A^\mu A^\nu{\cal R}_{\mu\nu}$ coupling sector: relative to GR, both the mass and the radius are smaller at low central densities, but become larger at high central densities. A comparison between the two coupling sectors further shows that their deviations are almost indistinguishable in the low-density regime, whereas clear differences emerge as the central density increases. In particular, the $A^2{\cal R}$ coupling leads to a faster growth of the stellar mass with $\rho_0$, so that the maximum-mass configuration appears at a lower central density. 
We also include in Fig.~\ref{fig:MRrelation} the observational mass constraint from the massive neutron star in a neutron-star--white-dwarf binary system, $M=2.01\pm0.04M_\odot$~\cite{Antoniadis:2013pzd}, together with the radius constraint $R_{1.4}=11.0^{+0.9}_{-0.6}\,{\rm km}$ for a $1.4M_\odot$ neutron star inferred from GW170817 multimessenger observations and nuclear-physics input~\cite{Capano:2019eae}. The representative parameter choices considered here are broadly compatible with these observational constraints.

\begin{figure}[]
\includegraphics[width=0.95\linewidth]{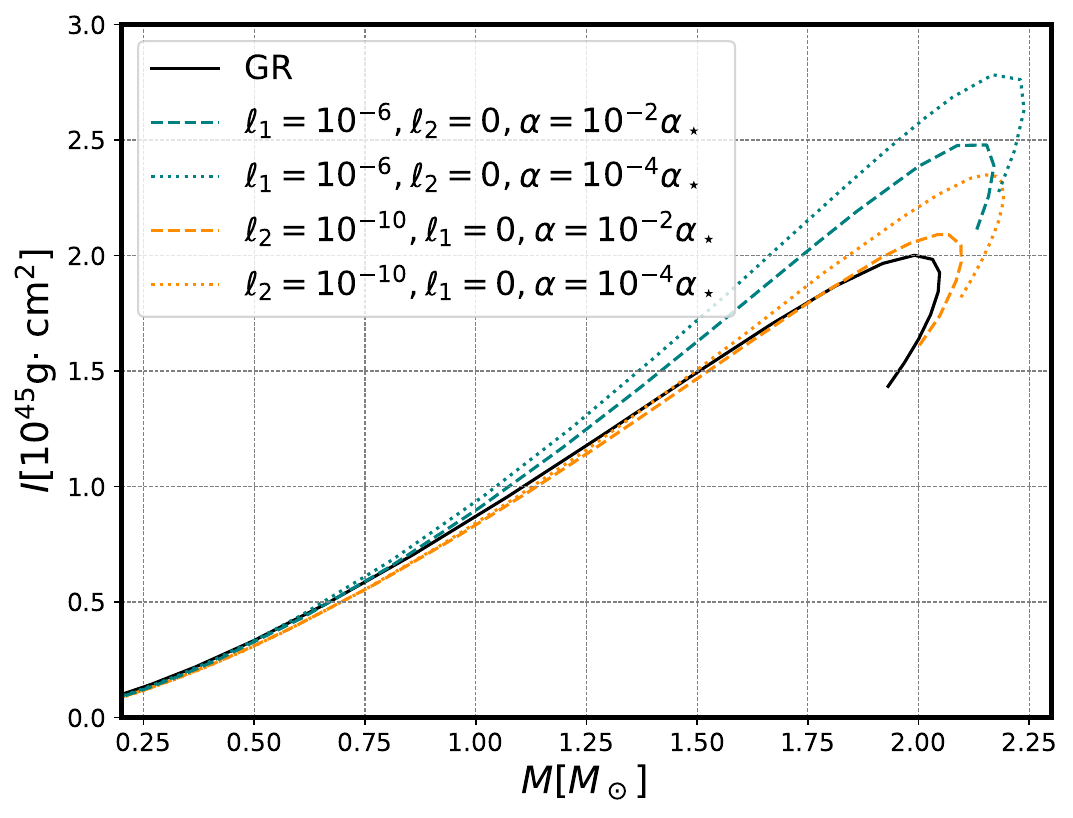}
\caption{\label{fig:IMrelation}Moment of inertia-mass ($I$–$M$) relations for neutron stars described by the SLy EOS in massive Hellings-Nordtvedt theory. We compare the two single-coupling sectors: the $A^2{\cal R}$ sector with $(\ell_1,\ell_2)=(10^{-6},0)$ and the $A^\mu A^\nu{\cal R}_{\mu\nu}$ sector with $(\ell_1,\ell_2)=(0,10^{-10})$. The data for the latter sector are taken from Ref.~\cite{Luo:2026oxw}. }
\end{figure}

Finally, we discuss the effects of the nonminimal coupling and the potential on the rotational properties of neutron stars. In the slow-rotation approximation, these effects are encoded in the moment of inertia $I$, defined in Eq.~\eqref{moi}. The $I$-$M$ relations are shown in Fig.~\ref{fig:IMrelation}.
Relative to GR, the $A^2{\cal R}$ coupling decreases the moment of inertia for low-mass stars but increases it for high-mass stars. This behavior is consistent with the trend observed in the mass-radius relation. Since the moment of inertia is sensitive to both the stellar mass and the radial distribution of matter, the reduction of the mass and radius in the low-central-density regime leads to a smaller $I$, whereas their enhancement in the high-central-density regime leads to a larger $I$. A comparison with the $A^\mu A^\nu{\cal R}_{\mu\nu}$ sector shows that the two couplings lead to similar qualitative behavior, but the quantitative difference becomes more visible toward the high-mass branch. The change in $I$ therefore provides another indication that the $A^2{\cal R}$ coupling can produce appreciable strong-field effects even for the weak-field-compatible value $\ell_1=10^{-6}$. Overall, these results indicate that massive nonminimally coupled vector-tensor gravity exhibits a pattern qualitatively similar to scalar-tensor theories with nonminimal couplings and potentials~\cite{Yazadjiev:2014cza,Liu:2024wvw,Li:2025gna,Yazadjiev:2016pcb,Staykov:2018hhc,Staykov:2014mwa}: the coupling and the potential jointly control the stellar structure, and sizable deviations can still arise in the strong-field regime even when the weak-field bounds are satisfied.

\section{Conclusion} \label{conclusion}

Vector-tensor gravity with the nonminimal coupling $A^\mu A^\nu{\cal R}_{\mu\nu}$ and a potential whose zero-energy minimum occurs at a nonzero value of $A^2$ is known to admit black-hole and neutron-star solutions with a global monopole-like asymptotic structure. In this work, we investigated whether this structure is generated solely by the potential and the associated nonzero vector vacuum, or whether it also depends on the specific form of the curvature-vector coupling. This question led us to study massive Hellings-Nordtvedt theory with the two independent nonminimal interactions $A^2{\cal R}$ and $A^\mu A^\nu{\cal R}_{\mu\nu}$.

We found that the global monopole-like asymptotic structure is coupling-dependent. The asymptotic vacuum branch selected by the potential, together with the field equations, does not allow the two nonminimal couplings to be simultaneously nonzero. As a result, the full theory is separated into two allowed single-coupling sectors. The sector with only the coupling $A^\mu A^\nu{\cal R}_{\mu\nu}$ reproduces the global monopole-like asymptotic structure found in previous studies, whereas the sector with only the coupling $A^2{\cal R}$ admits asymptotically flat vacuum solutions. Therefore, the global monopole-like asymptotics is not a generic consequence of the bumblebee-type potential or of the nonzero vector vacuum alone; rather, it is tied to the specific Ricci-tensor coupling.

We further investigated black holes and neutron stars in the $A^2{\cal R}$ sector. Although the black-hole metric in this sector takes the Schwarzschild form when written in terms of the integration constant, the Noether mass evaluated using the Wald covariant phase-space formalism differs nontrivially from this constant. This correction should not be interpreted as a universal feature of stealth black holes in vector-tensor gravity. It arises here because the nonminimal $A^2{\cal R}$ coupling, together with the nonzero vector vacuum selected by the potential, modifies the relation between the metric integration constant and the conserved charge. Once the solution is parametrized by the physical mass, the apparent degeneracy with GR is broken. This allows the Lorentz-violating parameter $\ell_1$ to be constrained by weak-field Solar-System tests. The resulting bounds constrain $\ell_1$ roughly within the range $-10^{-5}\lesssim \ell_1 \lesssim 10^{-6}$, with the strongest upper bound being of order $10^{-6}$.

Using a representative weak-field-compatible value $10^{-6}$ of $\ell_1$, we then constructed slowly rotating neutron-star solutions in the $A^2{\cal R}$ sector and studied their masses, radii, and moments of inertia. We found that, despite the small value of $\ell_1$ required by Solar-System tests, the $A^2{\cal R}$ coupling can still generate appreciable deviations from the GR predictions in the strong-field regime. The qualitative behavior is similar to that found in the $A^\mu A^\nu{\cal R}_{\mu\nu}$ sector: the mass, radius, and moment of inertia are reduced relative to GR in the low-central-density regime, but enhanced in the high-central-density regime. A direct comparison between the two sectors shows that their predictions are almost indistinguishable at low densities, whereas clear differences emerge along the high-density and high-mass branch.

The sizable deviations of neutron-star observables from their GR counterparts, despite the small deviations in the black-hole and weak-field regimes, further support the role of neutron stars as sensitive probes of strong-field gravity. At the level of the mass-radius relation, the representative parameter choices considered here are broadly compatible with the observational mass constraint from a massive neutron star and with the radius constraint for a $1.4M_\odot$ neutron star inferred from GW170817. It is therefore natural to extend the present analysis to additional neutron-star observables, such as the quadrupole moment and tidal Love numbers. These quantities would also make it possible to examine approximately EOS-insensitive universal relations, including the I-Love-Q relations~\cite{Yagi:2013bca,Yagi:2013awa}, and could provide further constraints on the Lorentz-violating parameters as well as on the form and parameters of the vector-field potential.

A more urgent issue, however, is the stability of the solutions. For the $A^\mu A^\nu{\cal R}_{\mu\nu}$ coupling without a potential, Ref.~\cite{Mai:2024lgk} studied black-hole perturbations and found that stable solutions can exist in certain regions of the parameter space. Ref.~\cite{Rubio:2024ryv} also discussed possible routes to avoiding dynamical pathologies associated with self-interacting vector fields~\cite{Coates:2022qia,Barausse:2022rvg,Coates:2022nif,Coates:2023dmz,Unluturk:2023qgk,Coates:2023swo,Unluturk:2024ltf,Rubio:2024ryv}. These results are encouraging, but they do not directly establish the stability of the black-hole and neutron-star solutions obtained in the present massive Hellings-Nordtvedt theory. Establishing their stability is therefore an important and urgent direction for future work.

\begin{acknowledgments}

We are grateful to Zexin Hu, Ziyi Li, Zhan-Feng Mai,  Zhi Xiao and Fangkang Yu for useful discussions. 
S.L. and H.Y. were supported in part by the National Natural Science Foundation of China (No. 12105098, No. 12481540179, No. 12075084, No. 11690034, No. 11947216, and No. 12005059) and the Natural Science Foundation of Hunan Province (No. 2022JJ40264), and the innovative research group of Hunan Province under Grant No. 2024JJ1006, and by the Excellent Young Scholars Program of the Hunan Provincial Department of Education under Grant No. 25B0092. 
H.D.L.~is  supported in part by Postdoctoral Innovation Project of Shandong Province SDCX-ZG-202503036 and National Natural Science Foundation of China Grants No.12447134. 
Z.X.Y is supported in part by the Young Scholars Startup Fund of Jining Normal University.

\end{acknowledgments}


\end{document}